\documentclass[12pt]{article}
\usepackage{epsfig}
\usepackage{amsmath}
\usepackage{longtable}

\usepackage{color}

\def\mua{\marginpar[\boldmath\hfil$\uparrow$]%
                   {\boldmath$\uparrow$\hfil}%
                    \typeout{marginpar: $\uparrow$}\ignorespaces}
\def\mda{\marginpar[\boldmath\hfil$\downarrow$]%
                   {\boldmath$\downarrow$\hfil}%
                    \typeout{marginpar: $\downarrow$}\ignorespaces}

\begin{document}

\newcommand{\be}{\begin{equation}}
\newcommand{\ee}{\end{equation}}
\newcommand{\fracm}{\displaystyle \frac}

\newcommand{\W}{M_W^2}
\newcommand{\h}{M_{h^0}^2}
\newcommand{\HH}{M_{H^0}^2}
\newcommand{\A}{M_{A^0}^2}
\newcommand{\Hp}{M_{H^+}^2}
\newcommand{\lam}{\fracm{\lambda_5 \W}{g^2}}
\newcommand{\sbb}{s_{2\beta}}
\newcommand{\cbb}{c_{2\beta}}
\newcommand{\saa}{s_{2\alpha}}
\newcommand{\caa}{c_{2\alpha}}
\newcommand{\sab}{s_{\alpha + \beta}}
\newcommand{\cab}{c_{\alpha + \beta}}
\newcommand{\sba}{s_{\beta - \alpha}}
\newcommand{\cba}{c_{\beta - \alpha}}
\newcommand{\pa}{\left(2\cab + \sba \saa \right)}
\newcommand{\pb}{\left(2\sab - \cba \saa \right)}
\newcommand{\pc}{\left(2\sab - \cba \sbb \right)}
\newcommand{\pd}{\left(2\cab - \sba \sbb \right)}
\newcommand{\ca}{\fracm{g}{M_W \sbb}}
\newcommand{\cb}{\fracm{g}{2 M_W}}
\newcommand{\cc}{\fracm{g^2}{4 \W \sbb^2}}
\newcommand{\cd}{\fracm{g^2}{4 \W \sbb}}
\newcommand{\ce}{\fracm{g^2}{4 \W}}
\newcommand{\cf}{\fracm{g^2}{2 \W}}

\newcommand{\RE}{\color{red}}
\newcommand{\BLU}{\color{blue}}
\newcommand{\BLA}{\color{black}}

\parindent 1.3cm
\thispagestyle{empty}   
 \vspace*{-3cm}
\noindent

\def\arccot{\mathop{\rm arccot}\nolimits}
\def\sd{\strut\displaystyle}

\begin{obeylines}
\begin{flushright}
\quad
\end{flushright}
\end{obeylines}

\vspace{2cm}

\begin{center}
\begin{bf}
\noindent
Pair Production of Two-Higgs-Doublet Model
Light Higgs Bosons  \\[0.2cm] 
in {\boldmath{$\gamma \gamma$}} Collisions
\end{bf}
  \vspace{1.5cm}\\
Fernando Cornet
\vspace{0.1cm}\\
Departamento de F\'\i sica Te\'orica y del Cosmos and CAFPE,\\
Universidad de Granada, E-18071 Granada, Spain\\
\vspace{0.5cm}
Wolfgang Hollik
  \vspace{0.1cm}\\
Max-Planck-Institut f\"ur Physik \\
F\"ohringer Ring 6, D-80805, M\"unchen, Germany
   \vspace{2.2cm}

{\bf ABSTRACT}
\end{center}

We study the production of a pair of light, neutral, CP-even 
Higgs bosons in photon--photon  
collisions within the general Two Higgs Doublet Model (THDM). 
This is a process for which the lowest order contribution in both,
the Standard Model and the THDM, appears at one loop. We find that the
cross section for this process can be much larger in the THDM than in the 
Standard Model and the number of events expected at the Photon Collider
will allow a determination of some of the parameters in the scalar potential.

\newpage
The study of the electroweak symmetry breaking mechanism is one 
of the most important
topics for future collider experiments. The  
recent global fits to electroweak precision measurements suggest
that a light Higgs boson can be discovered in the near future at LHC 
(or possibly even at the Tevatron) \cite{EWdata}. Once such a particle  
is discovered the task 
will be to measure its properties and to investigate the level of
agreement with the Standard Model expectations, to either verify
the Standard Model Higgs mechanism or to necessitate 
the introduction of new physics concepts.
This task will
be better addressed at  future $e^+ e^-$ colliders.
Also, the Photon Collider can provide complementary
experimental data to study the properties of the Higgs boson.  

The simplest extension of the Standard Model Higgs sector is obtained by 
introducing a second Higgs doublet. In the most general case, 
such a THDM leads to unacceptable
large CP-violation and tree-level Flavor Changing Neutral Currents (FCNC). 
CP is conserved by the restriction to real parameters, and
tree level FCNC contributions are suppressed
by imposing a symmetry $\Phi_1 \to - \Phi_1$.
With these restrictions, there are
still two types of THDMs~\cite{Hall}. They differ in the way the Higgs 
doublets are coupled to the fermions. In Type-I, only one Higgs doublet
couples to the fermions, while in Type-II the neutral components of
the first Higgs doublet couple to up-type fermions and the neutral
components of the second Higgs doublet couple to down-type fermions. Indeed, 
the Higgs sector in the Minimal Supersymmetric Standard Model belongs to
this second type of a THDM. In both cases, after electroweak
symmetry breaking, the model is left with 5 scalar bosons: two neutral, 
CP-even bosons ($h^0$ and $H^0$), one neutral CP-odd boson ($A^0$), and
a pair of charged Higgs particles ($H^\pm$).

The phenomenology of an extended Higgs sector has been extensively studied
in the literature, but most of the work has been devoted to the Minimal
Supersymmetric Standard Model, where supersymmetry 
imposes significant restrictions to the structure of the scalar potential
\cite{Gunion,Heinemeyer}. Concerning the general THDM it has been shown that 
precise measurements of the decay 
widths $h^0 \to b \overline{b}$, $h^0 \to \gamma \gamma$, $h^0 \to
\gamma Z$ can provide crucial information on the scalar potential 
parameters~ \cite{Arhrib,Ginzburg}, as well as $B$ decays and 
electroweak precision data~\cite{Osland}.
Double Higgs boson production at the LHC can also be used
to probe deviations from the Standard Model value of the triple Higgs 
coupling \cite{Moretti}. Triple Higgs boson 
production processes in $e^+ e^-$
collisions at ILC appear as promising channels to study the
Higgs potential due to the large number of events expected \cite{Ferrera}.
The processes $e^+e^- \to \phi_i \phi_j Z$ with $\phi_i = h^0$, $H^0$, $A^0$
or $H^\pm$ \cite{Arhrib2} as well as the quantum corrections to
$e^+ e^- \to H^+ H^-$ \cite{Guasch} have also been shown to be of interest
in the determination of the Higgs self-coupling parameters.

In this paper we assume the CP-conserving THDM and
discuss the production of a pair of the lightest
CP-even Higgs boson $h^0$ in photon--photon collisions, 
$\gamma\gamma \to h^0 h^0$,
as expected at the Photon Collider. 
We are going to restrict our discussion to the
particularly interesting case where $h^0$ couples to gauge bosons 
and fermions as the Standard Model Higgs particle ($H$) does.
For such a situation, the experimental signatures of
$h^0$ are in general very similar 
to the ones of $H$ making the experimental distinction 
between both models a challenging task.
Since pair production of Higgs bosons in $\gamma\gamma$ processes is
loop-induced in both models, either standard and non-standard contributions
appear at the same level, thus advocating this process as a particular 
sensitive tool to probe the type of the Higgs sector.

The organization of the paper is the 
following. We will first recall the structure of the THDM 
and its free parameters, which define the couplings
that enter the amplitude for $\gamma\gamma \to h^0 h^0$.
Next, we will briefly discuss the Standard Model
predictions for the $\gamma \gamma \to H H$ cross section and number of 
events expected at the Photon Collider. This yields the
reference values that will be used 
for comparison with our results for THDM Higgs bosons,
which we will present thereafter, with emphasis on the mass of
the charged Higgs boson.
In our calculations we have used the packages
FormCalc and FeynArts \cite{Hahn}  

The most general potential for the extension of the scalar sector
of the Standard Model to include two  $SU(2)_L$ doublets, 
$\Phi_1$ and $\Phi_2$, with $Y=1$ reads as follows~\cite{Gunion},
\begin{equation}
\label{potential}
\begin{split}
V &  =  \lambda_1 \left( \Phi_1^\dagger \Phi_1 - v_1^2 \right)^2
      + \lambda_2 \left( \Phi_2^\dagger \Phi_2 - v_2^2 \right)^2
      + \lambda_3 \left[\left( \Phi_1^\dagger \Phi_1 - v_1^2 \right)
                       +\left( \Phi_2^\dagger \Phi_2 - v_2^2 \right) \right]^2 \\
\quad & + \lambda_4 \left[ \left( \Phi_1^\dagger \Phi_1 \right)
                           \left( \Phi_2^\dagger \Phi_2 \right) -
                           \left( \Phi_1^\dagger \Phi_2 \right)
                           \left( \Phi_2^\dagger \Phi_1 \right) \right] \\
\quad & + \lambda_5 \left[ Re \left( \Phi_1^\dagger \Phi_2 \right) - 
                                     v_1 v_2 \cos \xi \right]^2
        + \lambda_6 \left[ Im \left( \Phi_1^\dagger \Phi_2 \right) - 
                                     v_1 v_2 \sin \xi \right]^2 \\
\quad & + \lambda_7 \left[ Re \left( \Phi_1^\dagger \Phi_2 \right) - 
                                     v_1 v_2 \cos \xi \right]
                    \left[ Im \left( \Phi_1^\dagger \Phi_2 \right) - 
                                      v_1 v_2 \sin \xi \right].
\end{split}
\end{equation}
The corresponding Lagrangian violates CP unless we take  $\lambda_7 = \xi = 0$ and
the rest of the parameters as real. Except for
the term proportional to $\lambda_5$ this CP-conserving potential is 
symmetric under
the discrete transformation $\Phi_1 \to - \Phi_1$. This symmetry
cancels all the contributions to FCNC processes. The term proportional
to $\lambda_5$ breaks the symmetry only in a soft way via a dimension-two
term. So, one can allow $\lambda_5$ to be different from zero without
entering into conflict with the experimental data on FCNC processes. The
parameters $v_1$ and $v_2$ in Eq.~(\ref{potential}) are the vacuum
expectation values of the Higgs fields:
\begin{eqnarray}
\left< \Phi_1 \right> =  
  \left( \begin{array}{c} 0 \\ v_1 \end{array} \right) & \quad \quad &  
\left< \Phi_2 \right> = 
  \left( \begin{array}{c} 0 \\ v_2 \end{array} \right).
\end{eqnarray}
>From the experimental value of the $W$ boson mass we can fix the sum
$v_1^2+v_2^2$. In this way we are left with seven free
parameters: $\lambda_i$ with $i = 1,\dots,6$ and $\tan \beta = v_2/v_1$.

We denote the masses of the physical particle spectrum by
\mda
$M_{h^0}$, $M_{H^0}$, $M_{A^0}$, $M_{H^+}$; thereby, we choose $h^0$ as the 
lighter and $H^0$ as the heavier one of the CP-even neutral bosons. 
The particle masses 
can be written in terms of the parameter set
$\lambda_1, \dots, \lambda_6, \, \tan\beta$.
\mua
Alternatively, one can use a more easily
measurable set of parameters to fix the model: $M_{h^0}$, $M_{H^0}$,
$M_{A^0}$, $M_{H^+}$, $\lambda_5$, $\tan \beta$ and $\alpha$, where 
most of the 
parameters in the potential have been replaced by the masses of the
physical bosons and the mixing angle $\alpha$ between the two CP-even 
neutral fields. 
Just a single parameter, $\lambda_5$, is kept in this set as 
a remnant of the original couplings in Eq.~(\ref{potential}). 
The relations between both sets of parameters can be found in
Ref.~\cite{Akeroyd}. When translating from the first to the second set 
of parameters, one has to take into account the restrictions imposed
on the values of $\lambda_i$ from perturbative unitarity. 

\begin{figure}
\begin{center}
\epsfxsize=10cm
\epsfbox{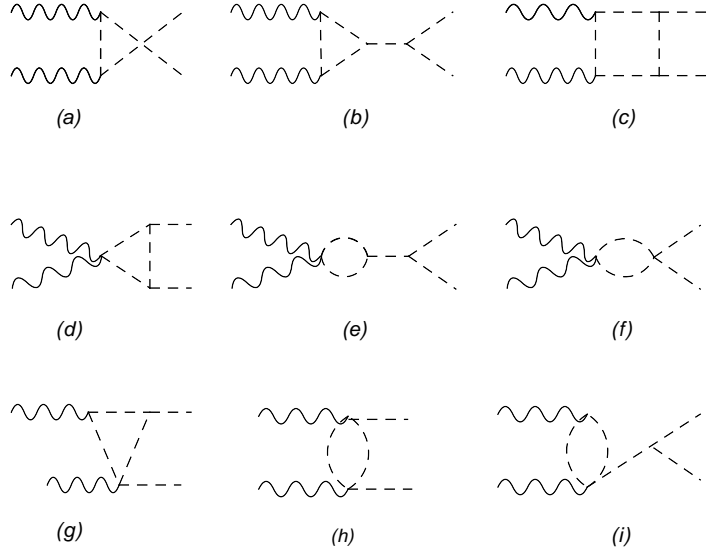}
\caption{\small  Diagrams contributing to neutral Higgs-boson pair 
production. In the 
Standard Model the particles in the loops in diagrams (a)-(d) can be all the
charged particles, while in diagrams (d)-(i) can only be $W$ bosons and
Goldstone bosons. In the THDM one can have in addition charged Higgs 
bosons in all loops and a heavy neutral Higgs in the $s$-channel
in diagrams (b),(e) and (i).}
\end{center}
\end{figure}

Before proceeding with the THDM, we set the reference scale 
and  briefly
review and update the results of Ref.~\cite{Jikia} for the 
Standard Model raction $\gamma \gamma \to HH$. 
Since the
Higgs boson does not directly couple to photons, 
the lowest order amplitude for
this process is of one loop order, 
just as in the case of single Higgs
production in two-photon collisions or the decay 
$H \to \gamma \gamma$~\cite{Ginzburg,Krawczyk}. 
The relevant set of diagrams are shown 
in Fig.~1 in a generic way. 
The particles running in the triangles and boxes (a)-(c) can be 
all the charged particles of the Standard Model, but the dominant 
contributions are obtained from the $t$ quark and $W$ boson. 
The diagrams
(d) to (i), however, only receive loop contributions from the $W$ boson
and Goldstone bosons.. 
It is interesting to point out that the triple Higgs vertex appears in the
diagrams (b),(e) and (i).

The production cross section 
for a Higgs boson  mass $m_H = 120$~GeV is shown in Fig.~2
as a function of the center-of-mass energy $E_{\gamma\gamma}$
in the relevant range for the Photon Collider.
Other parameter values are $m_t = 171.4$~GeV 
and $M_W = 80.40$~GeV. The red line corresponds to the 
configuration where both
photons have the same helicity, while the green line corresponds 
to the case where the two photons have opposite helicity. At these 
energies, near threshold, the configuration with the same helicity dominates
and shows some sensitivity to the top quark mass; indeed one can 
observe the effects of the $t \bar{t}$ threshold
in the change of slope in the red curve, but the dependence with the
$t$ quark mass is too small to be observable at the Photon Collider
for values of the mass within the present experimental error.
Convoluting these cross sections with the expected luminosities
at the Photon Collider \cite{Telnov} one can expect 39 events per year.

\begin{figure}
\begin{center}
\includegraphics[scale=0.49,angle=270]{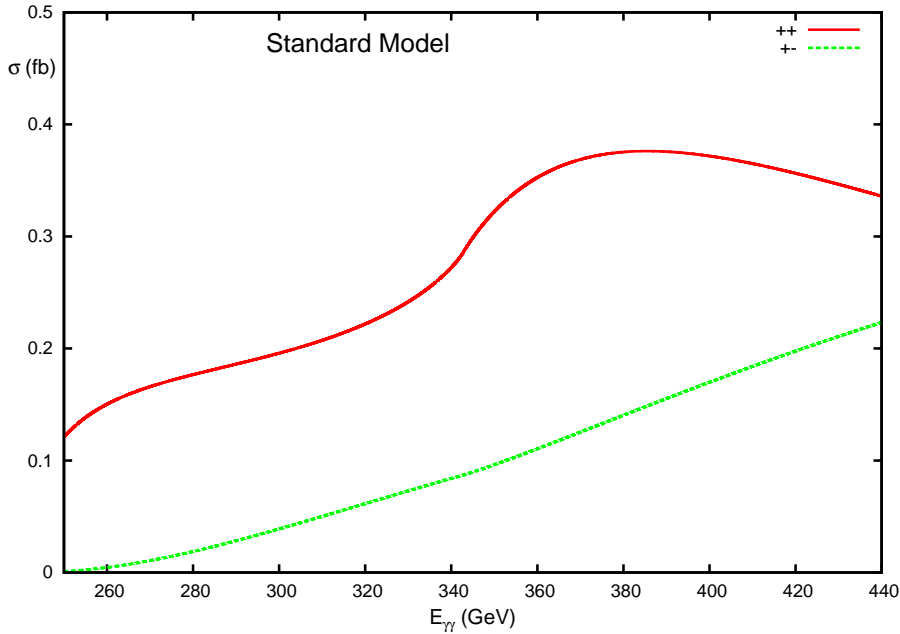}
\caption{\small Standard Model cross section as a function of the two-photon 
center-of-mass energy for the cases where the two photons have the 
same helicity (red line) and opposite helicity (green line).}
\end{center}
\end{figure}

The general CP-conserving Two Higgs Doublet Models 
introduce two types of modifications 
into the Standard Model calculation. First, the same set of generic diagrams 
contribute, but the couplings of the Higgs boson to the standard model 
particles and the triple Higgs $h^0h^0h^0$ coupling change. 
In particular, the couplings 
of the Higgs boson to the fermions depend on the type of THDM we consider. 
Second, new diagrams contribute to 
the process: (i) diagrams that have the
same form as the ones shown in Fig.~1, but the particles running in the loops 
are now charged Higgs bosons, and (ii) new diagrams similar to
(b) and (e) with a heavy neutral Higgs boson in the $s$-channel also 
contribute. Since the neutral CP-odd Higgs boson does not contribute
to this process, our results will be independent of the mass $M_{A^0}$.

In the following we restrict our discussion to the case where 
\be
\label{relab}
\alpha = \beta - \fracm{\pi}{2}.
\ee
This is a particularly interesting situation because the lightest Higgs boson
couples to the Standard Model Particles just in the same way as the standard
Higgs boson does. The differences between the Standard Model Higgs boson, $H$, 
and the lightest THDM neutral Higgs boson, $h^0$, only appear at
the  one-loop level.
Since the lowest order contribution to 
$\gamma \gamma \to hh$ is also of  one-loop order, 
this process becomes  most
appropriate to distinguish between the two models. 

Introducing the relation (\ref{relab}) for the THDM parameters has 
several consequences
First, since the coupling of $h^0$ to fermions is the same 
as for the standard $H$, 
it is independent of the type of the THDM. Hence
our results apply for both types of THDMs. 
Second, the triple coupling
$h^0 h^0 H^0$ vanishes, 
so no diagrams containing a $H^0$
contribute to our process. This makes our results 
independent of the mass of the heavy neutral Higgs boson. Third,
the only new contributions are those 
with charged Higgs bosons in the loops of Fig.~1.
The relevant couplings for these diagrams are: 
\be
\label{couplings}
\begin{array}{lcl}
h^0 H^+ H^- & \longrightarrow & - \fracm{ig}{2M_W} \left(\h + 2 \Hp 
                               - \fracm{4 \lambda_5 M_W^2}{g^2}\right), \\
\quad & \quad \\
h^0 h^0 H^+ H^- & \longrightarrow & -\fracm{ig^2}{4 \W} \left(\h + 2 \Hp 
                               - \fracm{4 \lambda_5 M_W^2}{g^2}\right) ,
\end{array}
\ee
with the $SU(2)$ gauge coupling $g$.
These self couplings turn out to be independent of $\tan \beta$.
In summary,
our results for the cross-section for $\gamma \gamma \to h^0 h^0$ 
depend only on three parameters: $M_{h^0}$, $M_{H^+}$, and $\lambda_5$.
Certainly, 
if we relax relation~(\ref{relab}) between
the angles $\alpha$ and $\beta$,
the cross section becomes dependent on the type of the THDM
and on all the parameters except $M_{A^0}$. But differences 
to the Standard Model will then appear already at the tree level.

First we consider the case of a ``light'' charged Higgs boson, which means 
that its mass is low enough for $H^\pm$ to be pair produced and, 
thus, discovered at a Linear Collider or the LHC. 
So, by the time of a Photon Collider,
the mass $M_{H^+}$ would be known and the only relevant
unknown parameter would be $\lambda_5$. 
Strictly speaking, this discussion is only meaningful 
for a THDM of Type I because 
in the THDM Type II one has a bound of
$M_{H^+} \ge 350$~GeV from $b \to s \gamma$ decays~\cite{Misiak}.
 This bound is valid for any value of $\tan \beta > 5$.

\begin{figure}[t]
\begin{center}
\includegraphics[scale=0.95,angle=270]{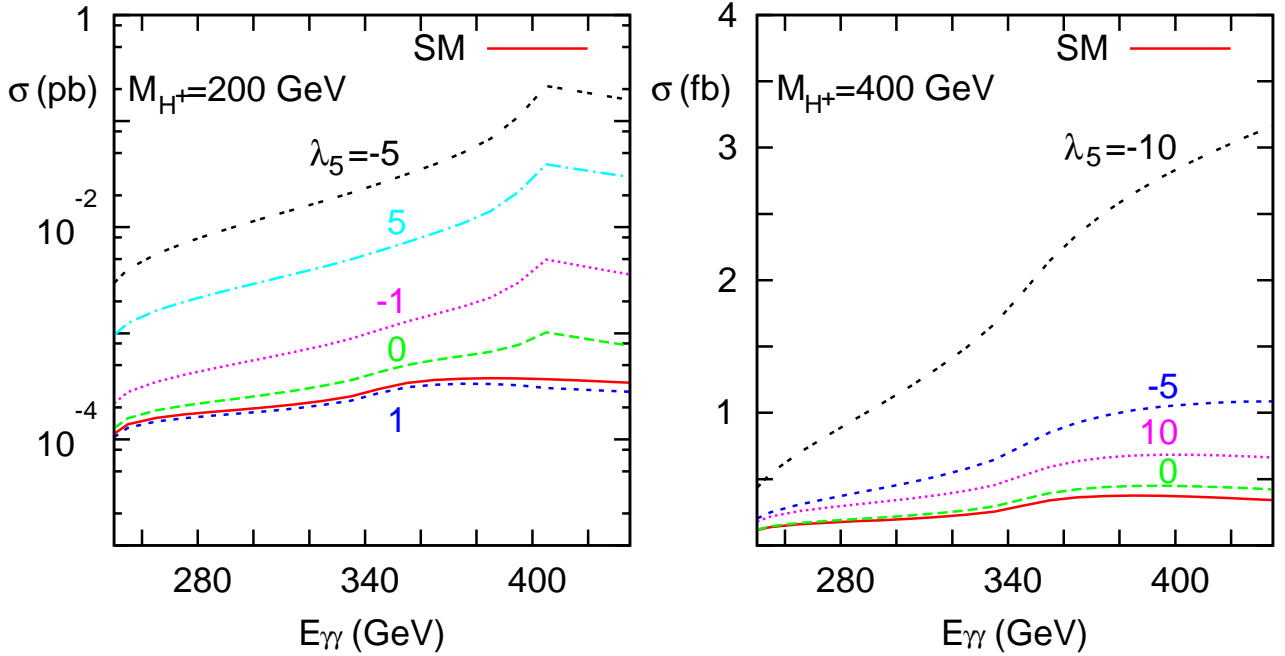}
\caption{\small Cross-section for $\gamma \gamma \to h^0 h^0$ for the helicity 
configuration in which both photons have the same helicity, as a function  
of the $\gamma \gamma$ center-of-mass energy}
\end{center}
\end{figure}

In the left plot of Fig.~3 we display the 
$\gamma \gamma \to h^0 h^0$ cross section as a function of the
center-of-mass energy $E_{\gamma\gamma}$ of the two photons
for the configuration where the  
photons have the same helicity, which yields the largest cross section. 
We have chosen, as an example, a mass $M_{H^+} = 200$~GeV and a range 
of values of $\lambda_5$ from $\lambda_5 = -5$ to $\lambda_5 = 5$
(well within the allowed range). The 
cross section turns out to be very sensitive to the value of $\lambda_5$,
particularly for negative values. One obtains an increase as large as three 
orders of magnitude for $\lambda_5= -5$. This is not very surprising because
the vertex $H^+H^-h^0$ in Eq.~(\ref{couplings}) appears twice in the
diagrams in Fig.~1, i.e. $\lambda_5$ appears squared in the amplitude. 

There are also values of the parameters for which the effects are very small. 
Indeed, from Eq.~(\ref{couplings}) it is clear that the triple and quartic 
couplings $H^+H^-h^0$ and $H^+H^-h^0h^0$ vanish when the relation
\begin{equation}
\label{neutral}
 2 M_{H^+}^2 + H_{h^0}^2 - \frac{M_W^2 s_W^2}{\pi \alpha} \lambda_5 = 0
\end{equation}
holds, thus reducing the differences between the Standard Model predictions
and the THDM predictions to the effects of the diagrams containing 
charged Higgses and  
Goldstone bosons in the loops.

The differences between the cross sections in the THDM and the Standard Model 
are reduced when
the mass of the charged Higgs boson is increased. 
We show, as an example, the 
cross section for the case of $M_{H^+} = 400$~GeV in Fig.~3, right plot. 
Still, even for such a higher charged Higgs mass, the effects can be 
observable for a wide range  of $\lambda_5$.

\begin{figure}[htb]
\includegraphics[scale=0.60,angle=270]{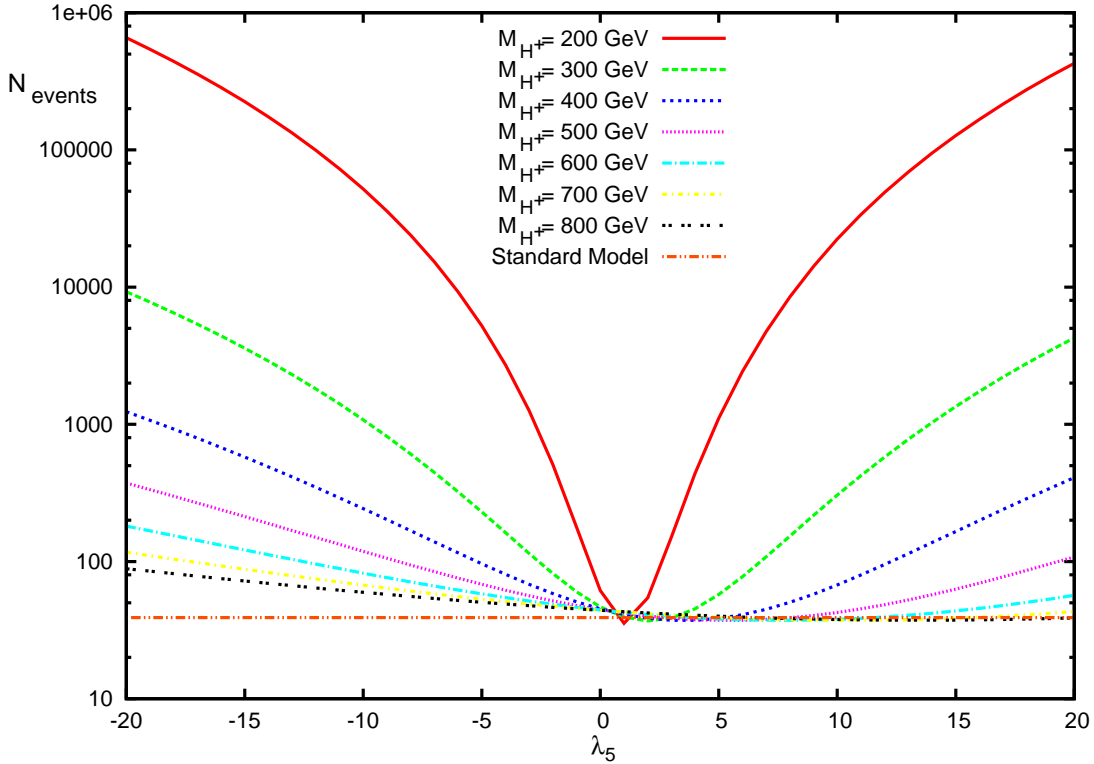}
\caption{\small Total number of events in the THDM as a function of 
$\lambda_5$ for different
values of the charged Higgs boson mass. 
The horizontal line represents the Standard Model 
prediction for the number of events}
\end{figure}

The predicted number of $h^0h^0$ pairs for the Photon Collider
as a function of $\lambda_5$ is shown in Fig.~4 for different values of
$M_{H^+}$, summing over all possible helicity configurations.
We have used the luminosities of Ref.~\cite{Telnov}. 
It is interesting to observe that for negative values of 
$\lambda_5$ the difference between the SM and the THDM predictions is 
large even for rather large values of the charged Higgs mass, whereas
for positive values of $\lambda_5$ the THDM cross section approaches
the SM result rather quickly, 
in such a way that for $M_{H^+}$ larger than about
700~GeV it will be very difficult to differentiate between the models.
The reason is that the relation (\ref{neutral}) cannot be satisfied
for $\lambda_5 < 0$, thus keeping the $H^+H^-h^0$ and $H^+H^-h^0h^0$ 
couplings always rather large, 
while for $\lambda_5 > 0 $ these couplings
become much smaller by partial compensation of the two terms..

To summarize, we have studied the process 
$\gamma \gamma \to h^0 h^0$, where $h^0$ is the lightest, neutral, CP-even
Higgs boson in the general Two Higgs Doublet Model. 
We have focussed our study to the intricate case 
where the $h^0$ couplings to the standard particles are the same
as the couplings of the Standard Model Higgs boson.
This means that tree level 
cross sections are the same in both models 
and differences appear only at higher order.
Since pair production of
neutral Higgs bosons in two-photon collisions is loop-induced, 
this is a very suitable place to look for differences between the SM and
THDM predictions. 
We have found that for a wide range of values
of $M_{H^+}$ and $\lambda_5$ 
(and independent of the residual model parameters, 
i.e.\ $M_{H^0}$, $M_{A^0}$, and $\tan \beta$) the cross section in
the THDM is much larger than in the Standard Model. Using the
predicted two-photon luminosities for the Photon Collider, the 
expected number of events, for negative values of the parameter
$\lambda_5$ and values of the charged Higgs boson mass up to $\sim 800$~GeV,
should be large enough to distinguish between the models and
to allow a determination of these parameters.

\bigskip
We thank A.~Arhrib and S.~Penaranda for useful comments. 
This work was supported in part by the
European Community's Marie Curie Research Training 
Network under contract MRTN-CT-2006-035505 
``Tools and Precision Calculations for
Physics Discoveries at Colliders'' (HEPTOOLS). 
FC thanks for financial support from
Spanish MEC under contracts FPA2006-05294 and Consolider - Ingenio 2010
Program CPAN (CSD2007-00042) and also Junta de Andalucia under 
contract FQM-330 and FQM-437. WH thanks the Aspen Center for Physics 
where the paper was finalized.

\end{document}